\documentclass[letterpaper, 10 pt, conference]{ieeeconf}
\IEEEoverridecommandlockouts
\usepackage{cite}
\usepackage{amsmath,amssymb,amsfonts}
\usepackage{algorithmic}
\usepackage{algorithm}
\usepackage{graphicx}
\usepackage{textcomp}
\usepackage{xcolor}
\usepackage{subfigure}
\usepackage{CJK}
\usepackage{indentfirst}
\usepackage{upgreek}
\usepackage{tabu}
\usepackage{booktabs}
\usepackage{threeparttable}
\usepackage{tabularx}
\usepackage{multirow}
\usepackage{multicol}
\usepackage{arydshln}
\usepackage{makecell}
\usepackage{verbatim}
\usepackage{epstopdf}
\newcommand{\req}[1]{\eqref{#1}} 
\usepackage{booktabs}

\def\BibTeX{{\rm B\kern-.05em{\sc i\kern-.025em b}\kern-.08em
    T\kern-.1667em\lower.7ex\hbox{E}\kern-.125emX}}
\begin{document}

\title{A Lifting Approach to Learning-Based Self-Triggered Control with Gaussian Processes}

\author{{Wang Zhijun}
\and
Kazumune Hashimoto
\and
 Wataru Hashimoto
\and
Shigemasa Takai \thanks{The authors are with the Graduate School of Engineering, Osaka University (e-mail: wang@is.eei.eng.osaka-u.ac.jp, hashimoto@eei.eng.osaka-u.ac.jp, hashimoto@is.eei.eng.osaka-u.ac.jp,takai@eei.eng.osaka-u.ac.jp)
}
}
\maketitle

\begin{abstract}
This paper investigates the design of self-triggered control for networked control systems (NCS), where the dynamics of the plant is {unknown} \textit{{apriori}}. 
To deal with the nature of the self-triggered control, in which state measurements are transmitted to the controller \textit{a}-periodically, 
we propose to lift the continuous-time dynamics to a novel dynamical model by taking an inter-event time as an additional input, and then, the lifted model is learned by the Gaussian processes (GP) regression. 
Moreover, we propose a learning-based approach, in which a self-triggered controller is learned by minimizing a cost function, such that it can take inter-sample behavior into account. By employing the lifting approach, we can utilize a gradient-based policy update as an efficient method to optimize both control and communication policies. 
Finally, we summarize the overall algorithm and provide a numerical simulation to illustrate the effectiveness of the proposed approach.
\end{abstract}


\section{Introduction}
In recent years, event-triggered and self-triggered control have attracted much attention 
since they are known to be useful strategies for saving resources in networked control systems (NCSs)\cite{survey}. 
In contrast to a time-triggered control that executes a feedback control law in a periodic manner, 
the event and self-triggered control transmit sensor measurements to the controller over the communication network only when it is needed. 
Various event/self-triggered controllers have been studied, see \cite{peng2018survey} for its survey paper. 
Early works apply the input-to-state stability or $L_2$ gain performance to design event/self-triggered control \cite{stability1} for linear systems. 
More recently, some approaches integrate the event/self-triggered control in optimal control\cite{optimalcontrol1,optimalcontrol2,optimalcontrol3}. 
In addition, researches related to reachability and safety analysis have been also provided in recent years\cite{safetyreachability1}. 

In many previous works of the event/self-triggered control framework, the dynamical model of the plant to be controlled is assumed to be {known} \textit{{apriori}}, 
which implies that when designed controllers are applied to real-world control systems, 
the achieved performance will heavily depend on the modeling accuracy. 
However, physical dynamics of the true system can be sometimes complex and highly nonlinear (and thus it is unknown \textit{apriori}). 
Motivated by this problem, several works have suggested an event/self-triggered control framework that can be applied for {unknown} transition dynamics; see \cite{learningbased,WANG2021203,deepev,FUNK2021100144,vam1,vam5,umlauft}. 
For example, \cite{learningbased} provided an iterative procedure of learning both the system dynamics and the optimal policy based on training data. In particular, they employed a Gaussian processes (GP) regression to learn the dynamics and a self-triggered optimal policy was obtained by solving a value iteration algorithm. 
Moreover, \cite{umlauft} investigates an approach to design an event-triggered controller for input-affine systems, where communication time instants are determined by evaluating a Lyapunov function candidate. 
In addition, \cite{deepev,FUNK2021100144,vam1,vam5,WANG2021203} investigated a model-free approach to designing event/self-triggered controllers based on a deep reinforcement learning framework. 

The objective of this paper is to learn a self-triggered controller based on an \textit{optimal control-based} framework, where the {unknown} transition dynamics is learned by the \textit{GP regression}. 
To the best of our knowledge, such an objective was achieved by only one previous work \cite{learningbased}.  However, we argue that this previous work has the following drawbacks. 
First, since the value iteration was implemented by \textit{discretizing} the state-space into grid points so as to approximate the optimal policy, it requires heavy computational resources in general. 
Second, \cite{learningbased} considers the dynamical system of the form: $x_{k+1} = f(x_k, u_k)$ ($x_k$ is the state and $u_k$ is the control input), and then the function $f$ is learned based on the training data while executing the self-triggered controller. However, when learning the function $f$ we require the training data as the \textit{consecutive} states and the control input (i.e., $x_k, x_{k+1}, u_k$), which implies that the training data is available only when the inter-event time step is $1$. This means that we have to throw away all the data (state/control inputs) if the inter-event time step is selected larger than 1, and thus it may not be of suitable for learning the dynamics while executing the self-triggered controller. 
Hence, the previous work has potential drawbacks in terms of both the requirement of computational resources and the inefficiency of data to learn the dynamics.

To address the issues as mentioned above, this paper proposes a novel optimal control, learning-based approach to designing the self-triggered controller with the GP regression. 
In particular, the proposed approach has the following contributions. 
First, in order to efficiently learn the dynamics and take the follow-up designing of the self-triggered controller into account, we choose \textit{not} to learn the dynamics originally defined as the ordinary differential or the difference equation (which is typically done in previous works of literature), but instead, we proposed to \textit{lift} to a model, in which an inter-event time is regarded as an additional control input, and then such a lifted model is learned by the GP regression. As will be detailed in later sections, this allows us to increase an efficiency of learning the dynamics, since we can utilize all the state information received at the controller as the training data to learn the dynamics. Second, we formulate a finite horizon optimal control problem, in which we can penalize the states between every adjacent triggering instants.
For example, this allows us to optimize an effective self-triggered controller that can avoid colliding with obstacles, as will be clarified in the numerical simulation. 
Finally, by employing the lifted dynamics estimated by the training data, we show that the self-triggered controller can be optimized via a policy gradient algorithm, which is one of the computationally efficient framework to derive the policy. This leads to a significant reduction of the computational time in contrast to the previous work \cite{learningbased}. 

As briefly mentioned above, our approach is also related to several previous works \cite{deepev,FUNK2021100144,vam1,vam5,WANG2021203,umlauft,key}, while the problem setup considered in this paper is significantly different from them in the following ways. \cite{deepev,FUNK2021100144,vam1,vam5,WANG2021203} investigate an approach to learn event/self-triggered controllers based on a deep reinforcement learning. 
Our approach differs from these works, in the sense that we here provide a \textit{model-based} solution, in which the (lifted) dynamics is learned by the GP regression and the optimal self-triggered controller is designed accordingly. 
Moreover, in constrast to \cite{umlauft} in which an event-triggered controller is designed based on a Lyapunov function candidate, our approach investigates a way to design a  \textit{self-triggered} controller based on an \textit{optimal control problem} with a policy gradient technique. 
In addition, our approach is related to \cite{key}, where an optimal control policy is designed via policy gradient for the dynamics learned by the GP regression. The proposed approach differs from \cite{key} in the following sense.
First, while \cite{key} investigates a way to design only a control policy, we here investigate a way to design a self-triggered controller that provides both control and communication policies, which will be achieved by introducing the lifted model. 
Second, we provide several modifications on a computational graph for policy evaluations and policy improvements. 
This is because we modify a cost such that the inter-sample behavior of the states can be taken into account while executing a self-triggered controller (for details, see Section~III-E,F).

\textit{Notation:}
Let $\mathbb{N}$, $\mathbb{N}_{>0}$ be the set of non-negative integers and positive integers respectively. 
Let $\mathbb{R}, \mathbb{R}_{\geq 0}, \mathbb{R}_{>0}$ be the set of reals, non-negative reals and positive reals, respectively. Given $x_{\min}, x_{\max} \in \mathbb{R}^n$, let $[x_{\min}, x_{\max}]$ denote a hyper-rectangle with the extreme (i.e., lower-left/upper-right) points $x_{\min}, x_{\max}$. 
For a square matrix $Q$, we use $Q \succ 0$ to denote that $Q$ is positive define. 
Let $\mathrm{diag}(a_1,a_2,\dots,a_h)$ be the diagonal matrix whose elements are give by $a_1,\dots,a_h \in \mathbb{R}$. Let $\mathbf{0}$ be the matrix of all zeros and $\mathrm{I}$ be the identity matrix.

\section{Problem Formulation}
\subsection{System description}
We consider the dynamical system of the form: 
\begin{equation}
     \label{model1}
     \dot{x}_{t} = f(x_{t}, u_{t}), \ u_{t} \in \mathcal{U},\ x_0 \sim \mathcal{N}(\mu_0, \Sigma_0),
\end{equation}
for all $t \in \mathbb{R}_{\geq 0}$, where $x_t \in \mathbb{R}^{n_x }$ and $u_{t} \in \mathbb{R}^{n_u}$ are 
the states and the control inputs at time $t$ respectively, and $\mathcal{U}= [u_{\min}, u_{\max}] \subset \mathbb{R}^{n_u}$ is the set of control inputs. We assume that the initial state $x_0$ follows the Gaussian with a given mean $\mu_0$ and a covariance matrix $\Sigma_0$. Moreover, $f : \mathbb{R}^{n_x } \times \mathbb{R}^{n_u} \rightarrow \mathbb{R}^{n_x }$ is a function representing the transition dynamics, which is {unknown} \textit{apriori}. 
\subsection{Overview of learning-based self-triggered control}

Let $t_0, t_1, t_2, \ldots $ with $t_0 = 0$ be the communication time instants when the plant transmits the state $x_{t_n}$ to the controller, 
and let $\tau_n = t_{n+1}-t_n \in \mathbb{R}_{>0}$, $n \in \mathbb{N}$ be the corresponding inter-event times. 
In this paper, we aim at designing a self-triggered controller so as to reduce the number of communication between the plant and the controller. 
The basic procedure of the self-triggered control is summarized as follows: for each $t_n$, $n\in\mathbb{N}$, 
\begin{enumerate}
\renewcommand{\theenumi}{(\roman{enumi}}
\setlength{\leftskip}{0cm}
\item The state ${x}_{t_n}$ is measured and transmitted to the controller; 
\item Based on some policy $\pi: \mathbb{R}^{n_x} \rightarrow \mathcal{U} \times \mathbb{R}_{>0}$, the controller computes the control input ${u}_{t_n}\in \mathcal{U}$ and the inter-event time $\tau_n \in \mathbb{R}_{>0}$, i.e., $[u_{t_n}^\top, \tau_n]^\top = \pi(x_{t_n})$;  
\item The controller transmits $\{u_{t_n}, \tau_n \}$ to the plant, and the plant applies $u_{t_n}$ constantly until the next communication time, i.e., $u_t = u_{t_n}$, $\forall t \in  [t_n, t_{n+1})$, where $t_{n+1} = t_n + \tau_n$. 
\end{enumerate}
Due to the fact that the dynamical system (\ref{model1}) is unknown \textit{apriori}, 
in this paper we employ a \textit{learning-based} approach, 
in which the controller learns the dynamical system based on training data and adaptively updates the policy $\pi$.
A rough sketch of the {learning-based} self-triggered control is summarized as follows: 
\begin{enumerate}
\renewcommand{\labelenumi}{[Step \arabic{enumi}]}
\setlength{\leftskip}{0.5cm}
\item Using the current policy $\pi$, implement the self-triggered control (i)--(iii) for a given time period $T \in \mathbb{R}_{>0}$. While executing the self-triggered controller, the controller stores a set of new training data involving the information of states received from the plant, control inputs and inter-event times. 
\item Using the training data, the controller learns the dynamical system and updates the policy $\pi$. Then, go back to Step~1.
\end{enumerate}

Our goal is to learn the optimal policy $\pi$ that minimizes a prescribed cost function (defined later in this paper). 
Moreover, we will ensure that the designed policy leads to satisfying $\tau_n \geq \tau_{\min}$, $\forall n \in \mathbb{N}$ for a given lower bound of the inter-event time $\tau_{\min} \in \mathbb{R}_{>0}$, which aims at guaranteeing the positive inter-event times. A concrete procedure of the proposed approach is elaborated in the next section. 

\section{Proposed Approach}
\subsection{Motivation} \label{S1}
A key challenge in learning-based self-triggered control given in the previous section is that the state measurements are transmitted to the controller \textit{only} at the communication time instants. 
Thus, the controller is able to utilize the states that are received {intermittently} from the plant, i.e., $x_{t_n}$, $n \in \mathbb{N}$, which indeed makes the learning of both the dynamics and the optimal policy a difficult task. 
As a naive approach, assuming that the lower bound of the inter-event time $\tau_{\min}$ is close to 0, 
one could often set the lowest inter-event time, say $\tau_n = \tau_{\min}$, 
so that the controller obtains the consecutive state measurements $x_{t_n}$, 
$x_{t_{n+1}}$ and use them to approximately learn the function $f(x_t, u_t)$, i.e., 
$\dot{x}_{t_n} = f(x_{t_n}, u_{t_n}) \approx (x_{t_n+\tau_n} - x_{t_n})/\tau_n$\footnote{Note that a direct access of $f(x, u)$, or the derivative of the state $\dot{x}$, is in general hard for digital computers. Hence, it is of practical to utilize the consecutive states $x_{t_n+\tau_n}, x_{t_n}$ in order to estimate $f$.}. 
Hence, if the controller receives the consecutive states $x_{t_n}$, 
$x_{t_{n+1}}$ with a small enough inter-event time, these states can be utilized as the training data to approximately learn $f$. 
However, this approach is clearly inefficient, 
since the controller is able to learn the dynamical system only for the case when the inter-event time is small enough, 
i.e., if the inter-event time is selected large, we need to possibly throw away the data since ($x_{t_n+\tau_n} - x_{t_n})/\tau_n$ is not accurate enough to estimate $f(x_{t}, u_{t})$. 
\textcolor{black}{In addition to the above, even if $f$ could be learned, deriving the optimal policy $\pi$ that minimizes a given cost function is in general computationally hard based on the knowledge about $f$. Indeed, the function $f$ depends on the control input $u$, but it does \textit{not} depend on the inter-event time $\tau$ (i.e., \req{model1} fails to incorporate the information about $\tau$), although we need to optimize both $u$ and $\tau$ for each state $x$. This implies that the standard policy gradient algorithm, which is known to be an efficient method to derive the optimal policy, cannot be directly applied based on the knowledge about $f$.} 

\subsection{{Lifting} approach for learning-based self-triggered control}
In order to efficiently and adaptively learn the dynamical system and the optimal policy while executing the self-triggered control, 
in this paper we propose to modify a \textit{function to be learned} as follows. First, note that we have 
\begin{equation}
        \label{model1Note}
        x_{t_{n+1}} = x_{t_n} + \int_{t_n}^{t_{n+1}} f(x_t, u_{t_n}) \mathrm{d}t, 
\end{equation}
where we use the fact that the control input is constant for all $t \in [t_n, t_{n+1})$. 
Then, letting the function $g$ be given by $g(x_{t_n}, u_{t_n}, \tau_n) = x_{t_n} + \int^{t_n + \tau_n} _{t_n} f(x_t, u_{t_n}) \mathrm{d}t$, 
we have
\begin{equation}
        \label{model2}
        x_{t_{n+1}} = g(x_{t_n}, v_n), \ n \in \mathbb{N}, 
\end{equation}
where $v_n = [u_{t_n}^\top,\tau_n]^\top$ denotes the \textit{extended} control input that incorporates both $u_{t_n}$ and $\tau_n$.
Instead of learning $f$, in this paper we propose to learn the \textit{lifted} function $g$ based on the training data. 
This approach is advantageous over learning $f$ in the following sense. 
First, in contrast to the case of learning $f$, whose training data is available only when the inter-event time is selected small enough as described above, we can make use of {all the state information} $x_{t_n}$, 
$n \in \mathbb{N}$ received at the controller as the training data to learn the dynamics; for details, see Section~III-C. 
Second, note that the inter-event time $\tau_n$ is now explicitly given as one of the inputs in $v_n$ and is incorporated in \req{model2}. As such, we can employ a policy gradient algorithm to compute the policy $\pi$, so that the optimal policy can be derived in a computationally efficient way; for details, see Section~III-F. 

\subsection{Learning the lifted dynamics with Gaussian Processes}

In this paper, we employ the GP regression in order to learn the \textit{lifted dynamics} \req{model2}. 
We independently predict each element of (\ref{model2}), which is denoted by $x_{t_{n+1},j}=g_j(x_{t_{n}},v_n),\ j = 1,2,\ldots,n_x$, where $x_{t_{n+1},j}$ and $g_j$ are the $j$-th element of $x_{t_{n+1}}$ and $g$, respectively. We denote by $\mathcal{D}_j = \{ {\tilde{X}}, Y_j\}$ the dataset to estimate $g_j$, where
\begin{align}
{\tilde{X}} &= \left [
\left [
\begin{array}{cc}
{x}^*_{t_{0}} \\
{v}^*_0
\end{array}
\right ], \left [
\begin{array}{cc}
{x}^*_{t_{1}} \\
{v}^*_1
\end{array}
\right], \ldots, 
\left [
\begin{array}{cc}
{x}^*_{t_{D-1}} \\
{v}^*_{D-1}
\end{array}
\right ]
\right ], \label{training_data1} \\ 
Y_j &= [\Delta^*_{0,j},\Delta^*_{1,j},\dots,\Delta^*_{D-1,j}]^\top, 
\label{training_data2}
\end{align}
with $\Delta_{n,j}^*=x_{t_{n+1},j}^* - x_{t_{n},j}^*$ for $n=0, \ldots, D-1$ and $D$ is the number of the training data. 
Then, given $\mathcal{D}_j$, an arbitrary state $x_{t_n} \in \mathbb{R}^{n_x}$ and an extended control input $v_n \in \mathbb{R}^{n_u+1}$, we can predict $x_{t_{n+1},j}$ by the Gaussian distribution as 
\begin{equation*}
        \begin{split}
                &p(x_{t_{n+1},j}\mid x_{t_n},v_{n},\mathcal{D}_j) = \mathcal{N}(x_{t_{n+1},j}\mid \mu_{n+1,j},\sigma_{n+1,j}),
        \end{split}
\end{equation*}
where $\mu_{n+1,j} = x_{t_n,j} + \mathbb{E}_g[\Delta_{n,j}], 
                \sigma_{n+1,j} = \mathrm{var}_g[\Delta_{n,j}]$
with $\Delta_{n,j}=x_{t_{n+1},j} - x_{t_n,j}$ are the mean and the variance of the GP prediction, respectively, and these are given by
\begin{equation}
        \label{gaussian2}
        \begin{split}
                \mathbb{E}_g[\Delta_{n,j}] &= {k}^T_{*,j}\left({K}_j+\sigma^2_{w,j} \mathrm{I}\right)^{-1} {Y}_j, \\
                \mathrm{var}_g[\Delta_{n,j}] &= k_j(\tilde{x}_{t_n},\tilde{x}_{t_n}) - {k}^T_{*,j}\left({K}_j+\sigma^2_{w,j} \mathrm{I}\right)^{-1}{k}_{*,j},
        \end{split}
\end{equation}
 where $\tilde{x}_{t_n}=[{x}_{t_n}^\top,v_n^\top]^\top$,  $k_j(\cdot,\cdot)$ is a given kernel parameterized by the hyperparameter $\theta_j$,  ${k}_{*,j} = k_j({\tilde{X}},\tilde{x}_{t_n}) \in \mathbb{R}^{D\times 1}$, and ${K}_j\in \mathbb{R}^{D\times D}$ is the kernel matrix whose $p$-$q$ element is defined as ${K}_{j,pq} = k_j(\tilde{x}^*_{t_{p}},\tilde{x}^*_{t_{q}})$, where $\tilde{x}^*_{t_n}=[x^{*\top}_{t_n}, v^{*\top}_n]^\top$. Moreover, $\sigma_{w}=\mathrm{diag}(\sigma_{w,1},\ldots,\sigma_{w,n_x })$ is a covariance of the Gaussian distributed white noise for the observation. 
After obtaining the predictions for all the elements of $x_{t_{n+1}}$, the whole predictive distribution of $x_{t_{n+1}}$ is $p(x_{t_{n+1}}\mid x_{t_n},v_{n},\{\mathcal{D}_j\}^{n_x}_{j=1}) = \mathcal{N}(x_{t_{n+1}}\mid \mu_{n+1},\Sigma_{n+1})$, where 
\begin{align}
    \mu_{n+1} &= [\mu_{n+1,1}, \ldots, \mu_{n+1,n_x }]^\top,\\
    \Sigma_{n+1} &= \mathrm{diag}(\sigma_{n+1,1},\ldots,\sigma_{n+1,n_x }). 
\end{align}
In other words, the obtained model provides, for given $x_{t_n}$ and $v_n = [u_n^\top, \tau_n]^\top$, a prediction of the state at the next communication time, i.e., $x_{t_{n+1}}$ with $t_{n+1} = t_n + \tau_n$. Note that, 
by learning this model and as shown in \req{training_data1} and \req{training_data2}, we can utilize \textit{all the information} of the states that are received from the plant, the control inputs, and the inter-event time as the training data to learn the dynamics. 

\subsection{Cost Function to Be Minimized}
\label{costfunction}

Let us now define the cost function to be minimized. 
Again, note that the model obtained in the previous section provides, for given $x_{t_n}$ and $v_n = [u_n^\top, \tau_n]^\top$, a prediction of the state at the next communication time $x_{t_{n+1}}$. 
Hence, as a simple and natural way for defining the cost function, one could consider the following: given $N \in \mathbb{N}$, 
\begin{equation}
    \label{naivecost}
    J(\pi_\psi)=\sum_{n=0}^{N-1} \mathbb{E}^{\pi_\psi}_{x_{t_{n}}} \left [ c(x_{t_n}, \tau_n)\right ], \ x_{t_0} \sim \mathcal{N}(\mu_0, \Sigma_0),
\end{equation}
where $\pi_\psi\ :\ \mathbb{R}^{n_x} \rightarrow \mathbb{R}^{n_u+1}$ denotes a policy parameterized by $\psi$, and $\mathbb{E}^{\pi_\psi}_{x_{t_{n}}} [\cdot]$ denotes the expectation with respect to $x_{t_{n}}$
conditioned on the policy $\pi_\psi$, and $c : \mathbb{R}^{n_x} \times \mathbb{R}_{\geq 0} \rightarrow \mathbb{R}_{\geq 0}$ denotes a given stage cost. \req{naivecost} implies that the stage cost is added at the communication time instants $t_n, n=0, \ldots, N-1$, which is in accordance with the transitions for the prediction model given in Section~III-C. 
Defining the cost function as above allows us to directly apply the policy gradient algorithm (see, e.g., \cite{key}) to optimize $\pi_{\psi}$. 
However, this cost function has a crucial drawback
that only triggered instants are considered, which means what happened between any adjacent triggered instants will not be taken into account. 
For instance, it could happen that the optimized policy makes the resulting state trajectory collide with an obstacle between the triggering instances; for details, see the simulation result in Section~IV. 
To overcome this shortcoming, we propose to modify the cost function as follows: given $M, N \in \mathbb{N}_{>0}$,  
\begin{align}
        \label{totalcost}
        \begin{split}
        J(\pi_\psi) =&\sum_{n=0}^{N-1} \mathbb{E}^{\pi_\psi}_{x_{t_{n,m}}}\left[\lambda c_1(\tau_n) +  \sum_{m=0}^{M-1} c_2(x_{t_{n,m}}) \right], \\ 
        & x_{t_0} \sim \mathcal{N}(\mu_0, \Sigma_0)
        \end{split}
\end{align}
where $t_{n,m} = \alpha_m t_{n+1}+(1-\alpha_m)t_n$ with
$\alpha_m = \frac{m}{M}$
for all $m=0, \ldots, M-1$. Moreover, $x_{t_{n,m}}$, $m=0, \ldots, M-1$, $n=0, \ldots, N-1$ is the state of the system by applying the control input $u_{t_n}$ constantly over the time length of $\alpha_m \tau_n$ from $x_{t_n}=x_{t_{n,0}}$, i.e., $x_{t_{n,m}}=g(x_{t_n}, v_{n,m})$ with $v_{n,m}=[u_{t_n}^\top,\alpha_m \tau_n]^\top$ and $[u^\top_{t_n}, \tau_n]^\top = \pi_\psi(x_{t_n})$. Moreover, $c_1$ and $c_2$ represent the stage cost for the communication and the state, respectively, and $\lambda>0$ is the weight associated to $c_1$. Examples of these cost functions include polynomials and mixtures of Gaussians, so that their expectations can be computed analytically (see Section~III-E). 
Intuitively, the cost related with control performance linearly \textit{interpolates} $M-1$ points between every adjacent triggering instants (i.e., $x_{t_{n,1}}, \ldots,  x_{t_{n,M-1}}$). 
As such, we can take the behavior of the states between every adjacent triggering instants into account. 

\subsection{Long-term prediction and policy evaluation}
To evaluate and minimize $J$ in (\ref{totalcost}), we need to predict the trajectory distribution
\begin{align}\label{trajectory}
    p(x_{t_{n,m}}),\ n=0,\dots,N-1,\ m=0,\dots,M-1
\end{align}
conditioned on the policy $\pi_\psi$. 
The calculation of (\ref{trajectory})
requires mapping a probability distribution through the GP model, which is mathematically intractable. 
Hence, we utilize a {moment matching} technique (see, e.g., \cite{key}) to approximate (\ref{trajectory}) by the Gaussians. 
While a basic procedure follows the approach given in \cite{key}, some modifications are necessary in terms of how to cascade the computations for $p(x_{n,m})$ (as detailed below). 
Here, we omit technical details of the moment matching technique, and provide only its summary and how the computational procedure is different from \cite{key}. 

\begin{figure}
        \centering
        \includegraphics[width=3.0in]{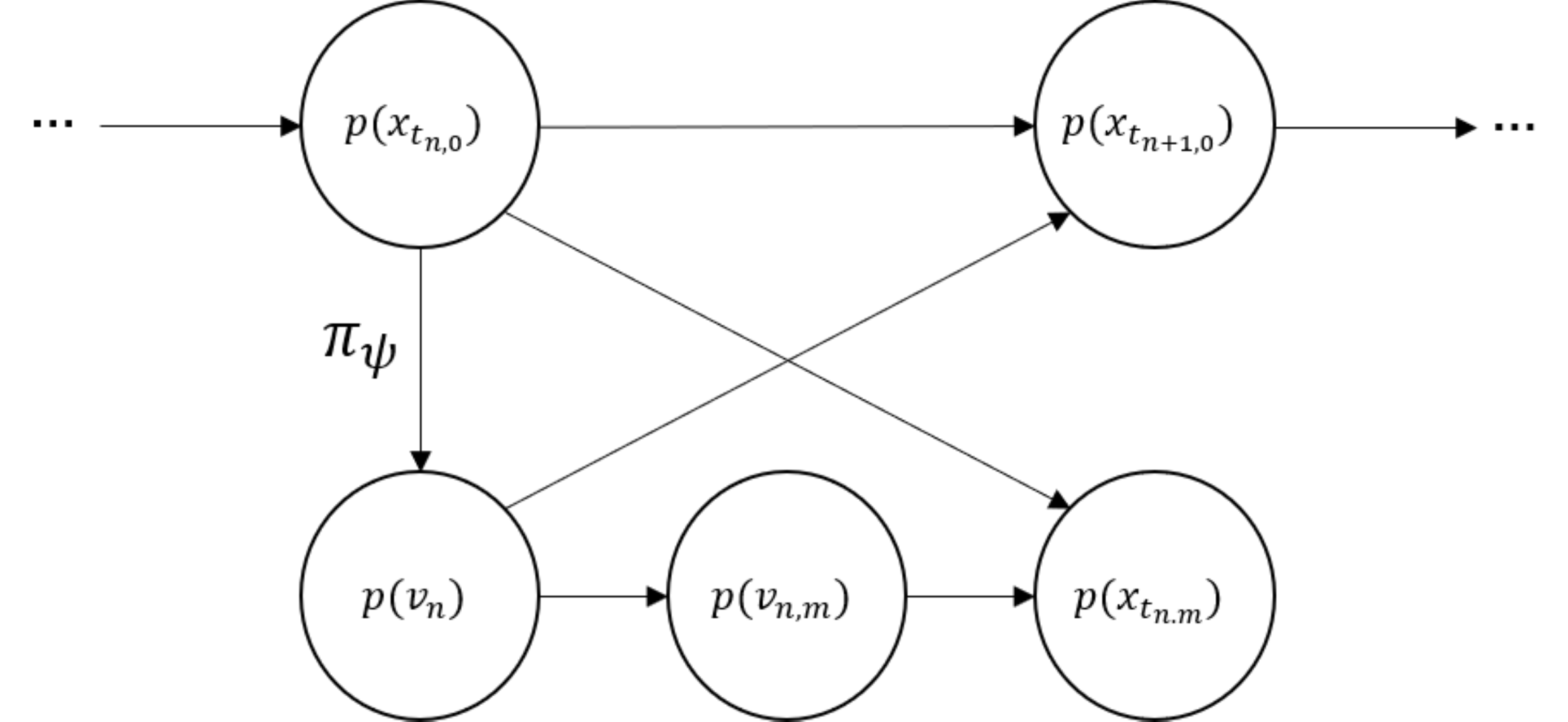}
        \caption{The computational graph of $p(x_{t_{n,m}})$ considered in this paper.}
        \label{comgr2}
\end{figure}

Suppose $p(x_{t_{n,0}})=p(x_{t_{n}})$ is approximated by a Gaussian distribution. 
Then, a Gaussian approximation for the distribution $p(v_{n}) = p(\pi_{\psi}(x_{t_{n,0}}))$ can be computed, and then we can analytically compute Gaussian approximation for the joint distribution $p(x_{t_{n,0}}, v_{n}) = p(x_{t_{n,0}}, \pi_{\psi}(x_{t_{n,0}}))$ (for details, see Section~5.5 in \cite{key}). 
Since $p(v_n)$ is Gaussian and $v_{n,m} (=[u_{t_n}^\top,\alpha_m \tau_n]^\top)$ is obtained from $v_{n}$ by the following linear transformation
\begin{equation}
        \label{U}
        \begin{split}  
        v_{n,m} & = \left[
                \begin{IEEEeqnarraybox*}[][c]{,c/c,}
                {\mathrm{I}}_{{n_u}\times{n_u}} & {0}_{{n_u}\times 1}\\
                {0}_{1\times {n_u}} & \alpha_m%
                \end{IEEEeqnarraybox*}
                \right]v_{n}, \\
        \end{split}
\end{equation}
  it follows that $p(v_{n,m})$ is also a Gaussian. 
  Since $p(x_{n,0})$ is Gaussian, we can analytically compute Gaussian approximation for the joint distribution $p(\tilde{x}_{t_{n,m}})$, where $\tilde{x}_{t_{n,m}}=[x_{t_{n,0}}^\top, v_{n,m}^\top]^\top$. 
Finally, the distribution of $x_{t_{n,m}}$ is computed from $p(x_{t_{n,0}}, v_{n,m})$ as follows: 
\begin{align}\label{eq:xtnm}
    p(x_{t_{n,m}})= \int p(x_{t_{n,m}}\mid\tilde{x}_{t_{n,m}}) p(\tilde{x}_{t_{n,m}}) \mathrm{d}\tilde{x}_{t_{n,m}}. 
\end{align}
From the GP model given in Section~III-C, it follows that $p(x_{t_{n,m}}\mid\tilde{x}_{t_{n,m}})$ is Gaussian. Therefore, by analytically computing the mean and covariance matrix of the right hand side of \req{eq:xtnm}, we can approximate the distribution of $x_{t_{n,m}}$ as a Gaussian distribution, i.e.,
$p(x_{t_{n,m}}) \approx \mathcal{N}(\mu_{n,m},\Sigma_{n,m})$. 
Similarly, we have 
\begin{align}\label{eq:xtnm2}
    p(x_{t_{n+1,0}})= \int p(x_{t_{n+1,0}}\mid\tilde{x}_{t_{n}}) p(\tilde{x}_{t_{n}}) \mathrm{d}\tilde{x}_{t_{n}},
\end{align}
where $\tilde{x}_{t_{n}} =[x_{t_{n,0}}^\top, v_{n}^\top]^\top$. From the GP model given in Section~III-C, it follows that $p(x_{t_{n+1,0}}\mid\tilde{x}_{t_{n}})$ is Gaussian.  
Hence, we can approximate $p(x_{t_{n+1,0}})$ by the Gaussian distribution by analytically computing the mean and the covariance in the right hand side of \req{eq:xtnm2}. 
Since we can iteratively cascade the above procedure, all of the required state distribution (\ref{trajectory}) can be obtained with this scheme. 
The computational graph that summarizes the above procedure
is shown in Fig.\ref{comgr2}. 

\begin{figure}
        \centering
        \includegraphics[width=2.5in]{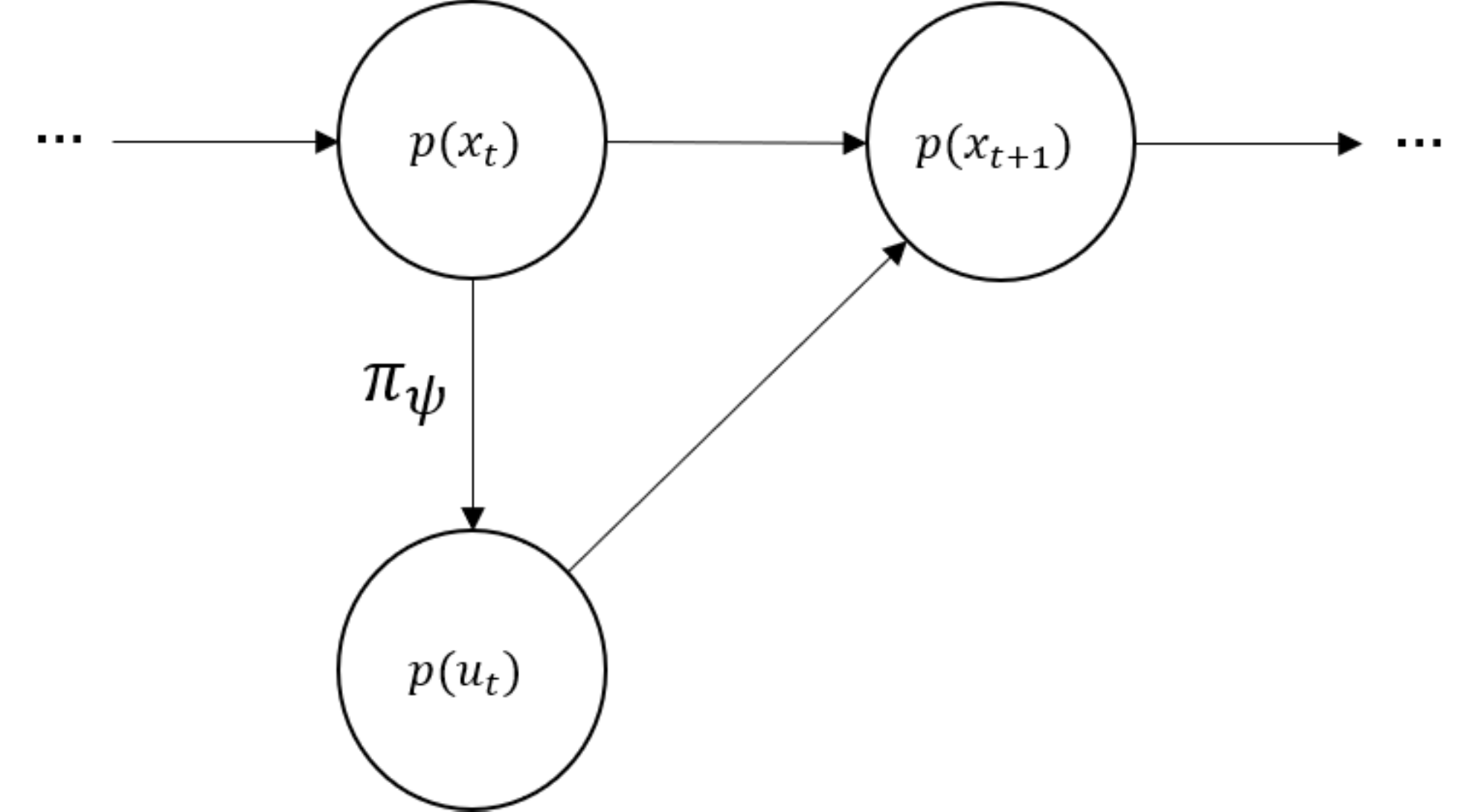}
        \caption{The computational graph originally proposed in \cite{key}.}
        
        \label{comgr3}
\end{figure}

For comparisons, in Fig.\ref{comgr3} we illustrate the computational graph originally proposed in \cite{key}. Note that \cite{key} considers a periodic control execution (i.e., the time interval between the time steps $t$ and $t+1$ in the above is always the same) and thus it aims at learning only a control policy. As shown in Fig.\ref{comgr3}, \cite{key} considers computing the state distribution iteratively based on the latest distributions of the state and the control input (i.e., $p(x_{t})$ is computed from $p(x_{t-1})$ and $p(u_{t-1})$). 
On the other hand, in our approach, the state distribution at the communication time instant is computed based on the distributions of the state and the extended control input at the latest communication time
(i.e., $p(x_{t_{n+1,0}})$ is computed from $p(x_{t_{n,0}})$ and $p(v_n)$). Moreover, all the state distributions \textit{between} the triggering instants, i.e., $p(x_{t_{n,m}})$, $m=1, \ldots, M-1$ are computed based on the distributions of the state at the latest communication time
$p(x_{t_{n,0}})$ and the extended control input at $t_{n,m}$, i.e., $p(v_{{n,m}})$ with $v_{n,m} = [u^\top_{t_n}, \alpha_m \tau_n]^\top$. 

Finally, to evaluate the expected the total cost $J$ in (\ref{totalcost}), 
it remains to compute the expected values
\begin{align*}
        &\mathbb{E}_{x_{t_{n,0}}}[c_1(\tau_n)]  = \int c_1(\tau_n) \mathcal{N}(\mu_{n,0},\Sigma_{n,0}) \mathrm{d}x_{t_{n,0}}\\ 
        &\mathbb{E}_{x_{t_{n,m}}}[c_2(x_{t_{n,m}})]  = \int c_2(x_{t_{n,m}}) \mathcal{N}(\mu_{n,m},\Sigma_{n,m}) \mathrm{d}x_{t_{n,m}}
\end{align*} 
Since $p(v_{n}) = p(\pi_{\psi}(x_{t_{n,0}}))$ is Gaussian, $p(\tau_{n})$ follows also a Gaussian. Hence, if the cost $c_1$ and $c_2$ are given by, e.g., polynomials, mixtures of Gaussians, we can analytically compute the above expectations. 
\subsection{Gradient Based Policy Improvement}
To find policy parameters $\psi$ minimizing $J(\pi_\psi)$ in (\ref{totalcost}), 
we use gradient information $\mathrm{d}J(\pi_\psi)/\mathrm{d}\psi$. 
As with the policy evaluation given in the previous section, we need to provide some modifications from \cite{key}, since the computational graph for (\ref{totalcost}) is different from that of \cite{key}. 
As a prerequisite, assume that the moments of the control distribution $\mu^v$ and $\Sigma^v$ can be computed analytically 
and are differentiable with respect to the policy parameters $\psi$. 
We obtain the gradient $\mathrm{d}J/\mathrm{d}\psi$ by repeatedly applying the chain rule. 
First, we use the notation:
\begin{equation}
        \begin{split}
                \zeta_{n} &= \lambda \eta_n +  \sum_{m=0}^{M-1} \epsilon_{n,m}, \\
                \eta_n &= \mathbb{E}_{x_{t_{n,0}}}[c_1(\tau_{n})],\  
                \epsilon_{n,m} = \mathbb{E}_{x_{t_{n,m}}}[c_2(x_{t_{n,m}})].
        \end{split}
\end{equation}
Then the following equations is obtained from (\ref{totalcost}). 
\begin{equation}
        \label{deri}
        \begin{split}
                \frac{\mathrm{d}J(\pi_\psi)}{\mathrm{d}\psi} &= \sum_{n=0}^{N-1} \frac{\mathrm{d}\zeta_n}{\mathrm{d}\psi}, \\
                \frac{\mathrm{d}\zeta_n}{\mathrm{d}\psi} &=\lambda \frac{\mathrm{d}\eta_n}{\mathrm{d}\psi}+ \sum_{m=0}^{M-1} \frac{\mathrm{d}\epsilon_{n,m}}{\mathrm{d}\psi}, \\
                \frac{\mathrm{d}\eta_n}{\mathrm{d}\psi} &= \frac{\mathrm{d}\eta_{n}}{\mathrm{d}p(\tau_n)} \frac{\mathrm{d}p(\tau_{n})}{\mathrm{d}\psi}, \\
                \frac{\mathrm{d}\epsilon_{n,m}}{\mathrm{d}\psi} &= \frac{\mathrm{d}\epsilon_{n,m}}{\mathrm{d}p(x_{t_{n,m}})} \frac{\mathrm{d}p(x_{t_{n,m}})}{\mathrm{d}\psi}.
        \end{split}
\end{equation}
We first discuss the term $\mathrm{d}\epsilon_{n,m}/\mathrm{d}\psi$ by adopting the shorthand notation $\mathrm{d}\epsilon_{n,m}/\mathrm{d}p(x_{t_{n,m}})=\left\{\mathrm{d}\epsilon_{n,m}/\mathrm{d}\mu_{n,m},\mathrm{d}\epsilon_{n,m}/\mathrm{d}\Sigma_{n,m} \right\}$ 
for taking the derivative of $\epsilon_{n,m}$ with respect to both mean and covariance of $p(x_{t_{n,m}})=\mathcal{N}(\mu_{n,m},\Sigma_{n,m})$ 
such that the following equation could be derived.
\begin{equation}
        \label{disderi}
       \frac{\mathrm{d}\epsilon_{n,m}}{\mathrm{d}\psi} = \frac{\mathrm{d}\epsilon_{n,m}}{\mathrm{d}\mu_{n,m}} \frac{\mathrm{d}\mu_{n,m}}{\mathrm{d}\psi} +\frac{\mathrm{d}\epsilon_{n,m}}{\mathrm{d}\Sigma_{n,m}} \frac{\mathrm{d}\Sigma_{n,m}}{\mathrm{d}\psi}.
\end{equation}
Next, the predicted mean $\mu_{n,m}$ and covariance $\Sigma_{n,m}$ 
depend on the moments of $p(x_{t_{n,0}})$ (see the computational graph in Fig.\ref{comgr2}) and the controller parameters $\psi$. 
By applying the chain rule to $\mathrm{d}p(x_{t_{n,m}})/ \mathrm{d}\psi$ 
in (\ref{deri}), we obtain
\begin{equation}
        \label{disderi2}
        \begin{split}
                &\frac{\mathrm{d}p(x_{t_{n,m}})}{\mathrm{d}\psi} = \frac{\partial p(x_{t_{n,m}})}{\partial p(x_{t_{n,0}})}\frac{\mathrm{d}p(x_{t_{n,0}})}{\mathrm{d}\psi}+\frac{\partial p(x_{t_{n,m}})}{\partial \psi},\\
                &\frac{\partial p(x_{t_{n,m}})}{\partial p(x_{t_{n,0}})} = \left\{ \frac{\partial \mu_{n,m}}{\partial p(x_{t_{n,0}})}, \frac{\partial \Sigma_{n,m}}{\partial p(x_{t_{n,0}})} \right\}.
        \end{split}
\end{equation}
From here onward, we focus on $\mathrm{d}\mu_{n,m} / \mathrm{d}\psi$, see (\ref{disderi}), 
because the calculation of $\mathrm{d}\Sigma_{n,m} / \mathrm{d}\psi$ is similar. 
For $\mathrm{d} \mu_{n,m} / \mathrm{d}\psi$, we calculate the derivative
\begin{equation}
        \begin{split}
                \frac{\mathrm{d}\mu_{n,m}}{\mathrm{d}\psi} &= \frac{\partial \mu_{n,m}}{\partial \mu_{n,0}} \frac{\mathrm{d}\mu_{n,0}}{\mathrm{d}\psi} 
               + \frac{\partial \mu_{n,m}}{\partial \Sigma_{n,0}} \frac{\mathrm{d}\Sigma_{n,0}}{\mathrm{d}\psi} 
               +\frac{\partial \mu_{n,m}}{\partial \psi}.
        \end{split}
\end{equation}
And $\mathrm{d}p(x_{t_{n,0}})/ \mathrm{d}\psi$ in (\ref{disderi2}) is known from communication time step $n-1$ since $x_{t_{n,0}}$ can be viewed as $x_{t_{n-1,M}}$. 
To compute $\mathrm{d}\mu_{n,m} / \mathrm{d}\psi$, it remains to compute 
\begin{equation}
        \label{disderiU}
        \begin{split}
                \frac{\partial \mu_{n,m}}{\partial \psi} &= \frac{\partial \mu_{n,m}}{\partial p(v_{n,m})} 
                                                        \frac{\partial p(v_{n,m})}{\partial \psi} \\
                &=\frac{\partial \mu_{n,m}}{\partial \mu^v_{n,m}} \frac{\partial \mu^v_{n,m}}{\partial \psi} + 
                \frac{\partial \mu_{n,m}}{\partial \Sigma^v_{n,m}} \frac{\partial \Sigma^v_{n,m}}{\partial \psi},
        \end{split}
\end{equation}
where $v_{n,m} \sim \mathcal{N}(\mu^v_{n,m},\Sigma^v_{n,m})$. 
The partial derivatives of $\mu^v_{n,m}$ and $\Sigma^v_{n,m}$, i.e. the mean and covariance of $p(v_{n,m})$, 
used in (\ref{disderiU}) depend on the policy representation. 
To evaluate $\mathrm{d}J(\pi_\psi)/\mathrm{d}\psi$, we also need the information of $\mathrm{d}p(\tau_n) / \mathrm{d}\psi$ in (\ref{deri}), 
which could be computed by 
\begin{equation}
        \label{disderifinal}
        \begin{split}
                \frac{\mathrm{d}p(\tau_n)}{\mathrm{d}\psi} &= \frac{\mathrm{d}p(\tau_n)}{\mathrm{d}p(v_{n})} \frac{\mathrm{d}p(v_{n})}{\mathrm{d}\psi} = 
                \left[
                        \begin{IEEEeqnarraybox*}[][c]{,c/c,}
                        \mathbf{0}_{1 \times {n_u}} & 1
                        \end{IEEEeqnarraybox*}
                \right]\frac{\mathrm{d}p(v_{n})}{\mathrm{d}\psi}, \\
                \frac{\mathrm{d}p(v_{n})}{\mathrm{d}\psi} &= \frac{\partial p(v_{n})}{\partial p(x_{t_{n,0}})}
                \frac{\mathrm{d}p(x_{t_{n,0}})}{\mathrm{d}\psi} + 
                \frac{\partial p(v_{n})}{\partial \psi},
        \end{split} 
\end{equation}
where $ \mathrm{d}p(x_{t_{n,0}}) / \mathrm{d}\psi$ and $\partial p(v_{n}) / \partial \psi$ has been discussed in the previous content, 
and $\partial p(v_{n}) /\partial p(x_{t_{n}})$ depends on the policy representation.
The individual partial derivatives in (\ref{deri}) to (\ref{disderifinal}) need to apply chain rule to moment matching (for their detailed computations, see the appendix of \cite{key}). 

\subsection{Guaranteeing positive inter-event times and control constraint satisfaction via parametrization}
One of the most desirable properties in self-triggered control is to guarantee a \textit{positive} inter-event time, i.e., there exists a $\tau_{\min}>0$ such that $\tau_n \geq \tau_{\min}$ for all $n = 0,1,\ldots $ (see, e.g., \cite{survey}). 
In addition, we constrain that the control input must belong to $\mathcal{U} = [u_{\min}, u_{\max}]$, i.e., $u_{t_n} \in \mathcal{U}$ for all $n = 0,1,\ldots $. In this section, we remark that these properties can indeed be satisfied by a parametrization technique. 

Let the policy $\pi_\psi$ be decomposed as $\pi_\psi(x) = \{\pi_{\psi_\tau}(x),\pi_{\psi_u}(x)\}$ 
where $\pi_{\psi_\tau}: \mathbb{R}^{n_x} \rightarrow \mathbb{R}_{\geq 0}$ denotes a policy to compute the inter-event time $\tau$, and $\pi_{\psi_u}(x): \mathbb{R}^{n_x} \rightarrow \mathbb{R}^{n_u}$ denotes a policy to compute the control input $u$. 
Now, let us parametrize the policy $\pi_{\psi_\tau}$ as follows: 
\begin{equation}
    \label{cons}
    \pi_{\psi_\tau}(x) = \frac{\tau_{\max}+\tau_{\min}}{2} + \frac{\tau_{\max}-\tau_{\min}}{2}\sigma_\tau(\tilde{\pi}_{\psi_\tau}(x)),
\end{equation}
for given $\tau_{\min}, \tau_{\max} >0$ \footnote{Here, $\tau_{\max}$ could be selected arbitrary large so as to lengthen the inter-event time.} with $\tau_{\max} > \tau_{\min}$ and $\sigma_\tau(\cdot)$ is a given squashing function that satisfies $\sigma_\tau(z) \in [-1, 1]$ for all $z \in \mathbb{R}$ (see Section~5.1 in \cite{key}), and $\tilde{\pi}_{\psi_\tau}: \mathbb{R}^{n_x} \rightarrow \mathbb{R}$ denotes a preliminary policy with an unconstrained amplitude, such as the one given by a radial basis function (RBF). 
Parametrizing $\pi_{\psi_\tau}$ as above leads to ensuring that $\pi_{\psi_\tau}(x) \in [\tau_{\min}, \tau_{\max}]$ for all $x \in \mathbb{R}^{n_x}$, and thus the optimized the policy ensures a positive inter-event time. Similarly, we can ensure a constraint satisfaction of the control input by parametrizing $\pi_{\psi_u}(x)$ as $\pi_{\psi^i _u}(x) = \frac{u_{\max,i}+u_{\min,i}}{2} + \frac{u_{\max,i}-u_{\min,i}}{2}\sigma_u(\tilde{\pi}_{\psi^i_u}(x))$ for all $i=1, \ldots, n_u$, where $u_{\max,i}$ and $u_{\min,i}$ denote the $i$-th element of $u_{\max}$ and $u_{\min}$ respectively, and $\pi_{\psi^i _u}$ denotes the $i$-th element of $\pi_{\psi_u}$ (i.e., it denotes a policy to compute the $i$-th element of $u$), $\sigma_u(\cdot) \in [-1, 1]$ is a given squashing function, and $\tilde{\pi}^i_{\psi_u}: \mathbb{R}^{n_x} \rightarrow \mathbb{R}$ denotes a preliminary policy with an unconstrained amplitude. 
Parametrizing $\pi_{\psi_u}$ as above leads to ensuring that $\pi_{\psi_u}(x) \in [u_{\min}, u_{\max}]$ for all $x \in \mathbb{R}^{n_x}$, guaranteeing the control constraint satisfaction. 
\subsection{Overall Algorithm}
Let us now introduce an overall implementation algorithm 
that jointly learns the dynamics of the plant and the self-triggered controller based on a model-based reinforcement learning framework. 
The overall algorithm is shown in Algorithm \ref{alg}. 
For the initial iteration, the controller generates random control signals (in a self-triggered manner) and apply to the system to record the training data in the form of $\{(x^*_{t_i},v^*_i),x^*_{t_{i+1}}\}$ according to Section~III-C (line~3). 
Then, using the recorded input and output data $\mathcal{D}$, the controller learns the lifted dynamics $x_{t_{n+1}} = g(x_{t_{n}}, v_n)$ 
using GP regression (line~5).
Next, the algorithm utilizes the current policy and the learned dynamics to predict future trajectory distribution 
$\{p(x_{t_{n,m}})\}$ from an initial state distribution $p(x_{t_{0}})$, and then calculates the expected total cost $J$ (line~7). 
Thereafter, the gradient information $\mathrm{d}J/\mathrm{d}\psi$ is computed and applied to minimize $J$, yielding an improved policy $\pi_\psi$. 
Then execute the control system based on the improved policy, and gather data $\{(x^*_{t_i},v^*_i),x^*_{t_{i+1}}\}$ during execution. 
The gathered data is appended to the total training data, and back to line 5, 
the dynamical model $x_{t_{n+1}} = g(x_{t_{n}}, v_n)$ is re-trained using the additional training data. 
Finally, the loop ends and outputs the optimal self-triggered control policy $\pi_\psi$ when a satisfactory performance is reached.

\begin{algorithm}
        \caption{Learning self-triggered controllers}
        \label{alg}
        \algsetup{linenosize = \small}
        \begin{algorithmic}[1]
        {\small
                \STATE \textbf{Input}: Characterization of the stage cost functions $c_1, c_2$, prediction horizon $N$, Gaussian distribution of the initial state $\mathcal{N}(\mu_0, \Sigma_0)$, initial policy parameter $\psi$, step-size of the gradient update $\alpha>0$;
               \STATE  \textbf{Output}: the optimal self-triggered control policy $\pi_\psi$;
               \smallskip 
		\STATE Starting from $x_0 \sim \mathcal{N}(\mu_0, \Sigma_0)$, apply the self-triggered controller for a given time period $T$, in which the controller generates a random extended control input $v_n$ for each communication time $t_n$. Then, record the training data (Section~III-C);  
		\REPEAT
		\STATE Using the recorded training data, estimate the lifted dynamics by the GP regression (Section~III-C);  
                \REPEAT
		\STATE Use the current policy $\pi_\psi$ to predict future trajectory distribution and calculate the expected total cost (\ref{totalcost}) (Section~III-E);
		\STATE Compute gradient information $\mathrm{d}J / \mathrm{d}\psi$ using (\ref{deri}) to (\ref{disderifinal});
                \STATE Update the policy parameter as $\psi \leftarrow \psi - \alpha \mathrm{d}J / \mathrm{d}\psi$; 
                \UNTIL $\psi$ converges
                \STATE Using the improved policy $\pi_\psi$, apply the self-triggered controller for a given time period $T$ and then record the training data;
		\UNTIL task learned
                }
        \end{algorithmic}
\end{algorithm}
\section{Simulation}
\subsection{Inverted pendulum}
\begin{figure}
        \centering
        \includegraphics[width=2.35in]{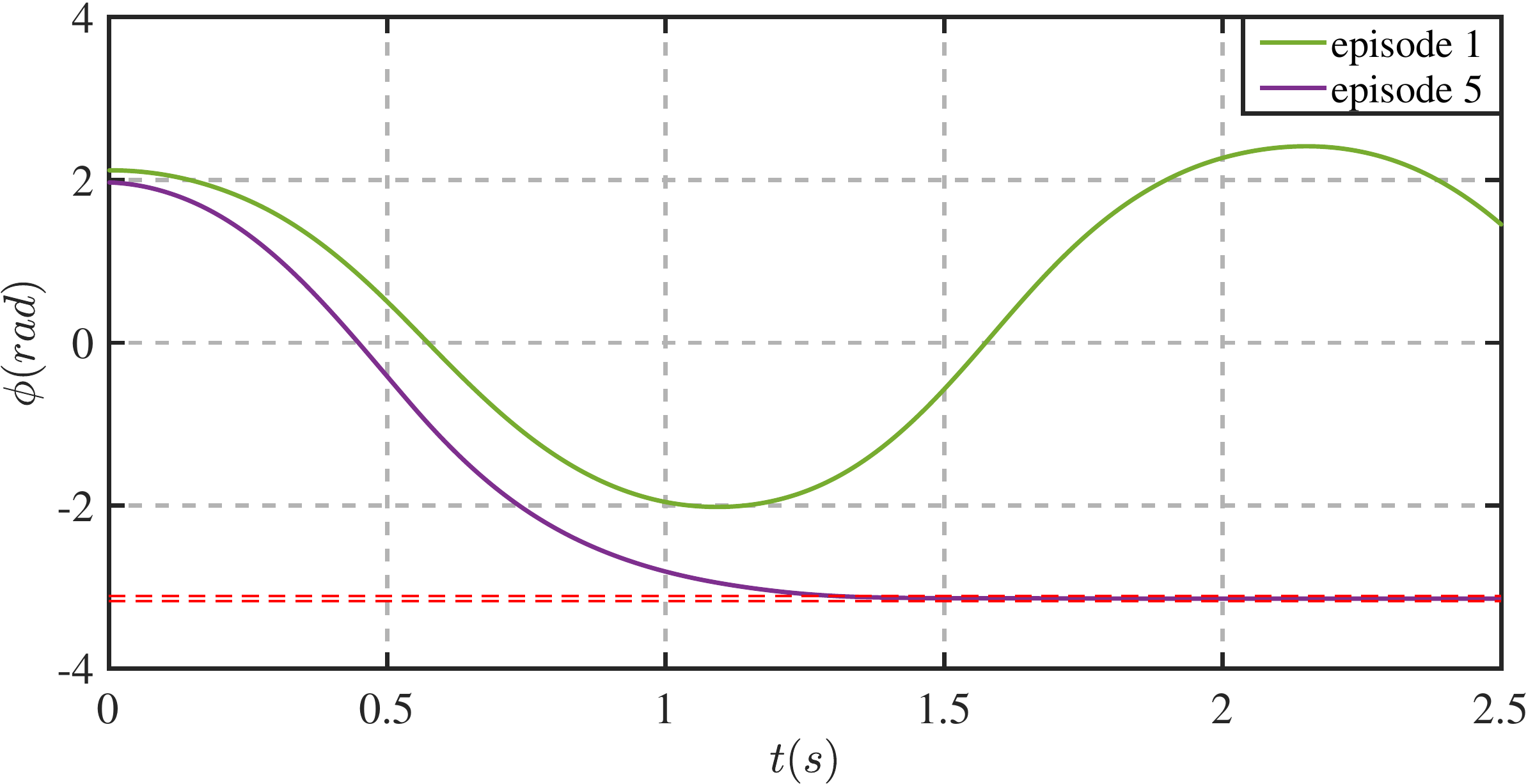}
        \caption{Learned Controller Performance when $\lambda =0.01$}
        \label{pendulum_Sim}
\end{figure}

\begin{figure}
        \centering
        \includegraphics[width=2.35in]{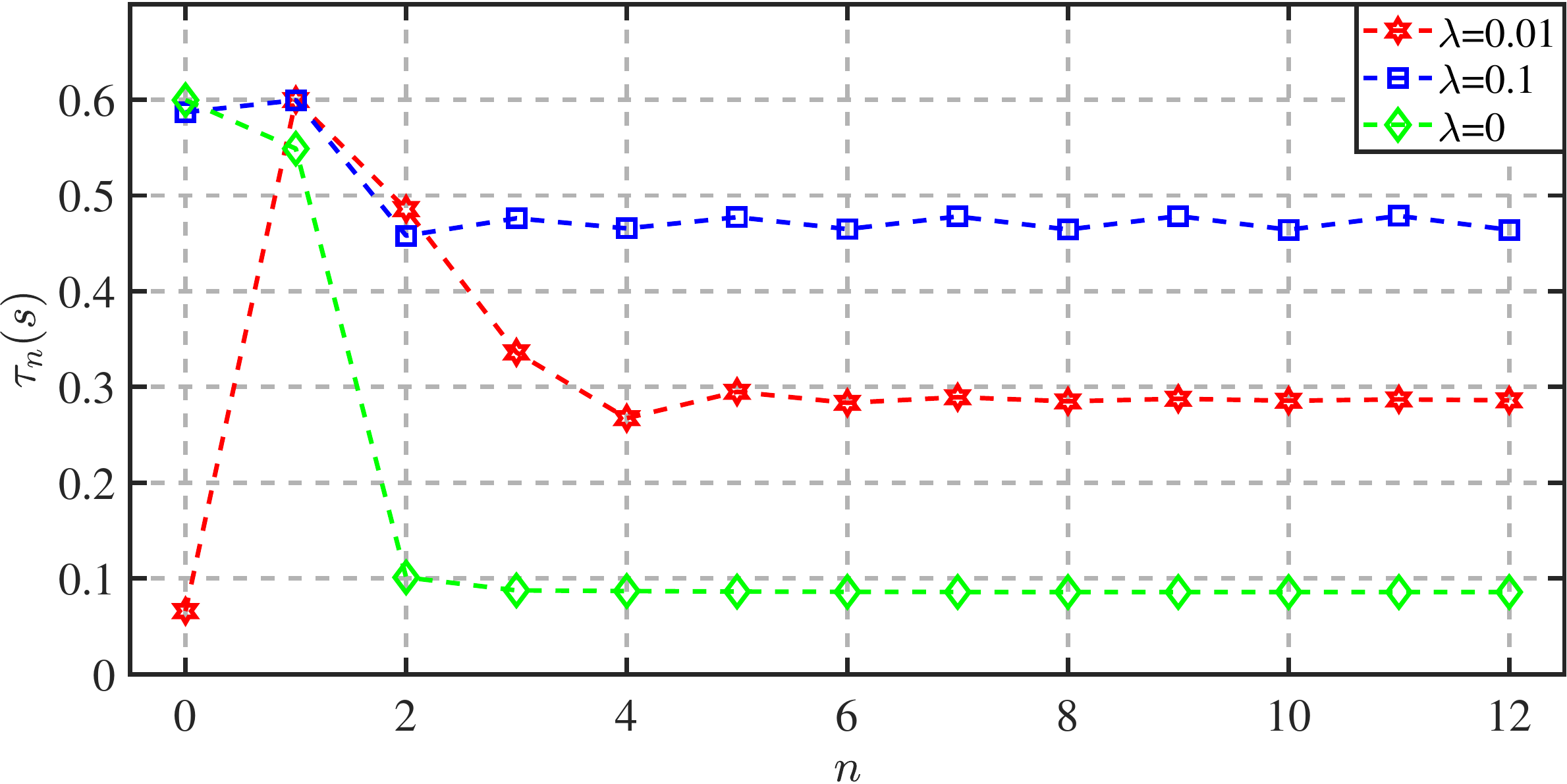}
        \caption{Inter-event time $\tau_n$ with $\lambda = 0.01,0.1,0$ when episode = $20$}
        \label{pendulum_tau}
\end{figure}

To make comparisons with the previous work \cite{learningbased}, we first conducted a simple experiment of an inverted pendulum, whose dynamics is given of the form: 
\begin{equation}
\label{pendulumDyn}
        \Ddot{\phi} = \frac{u-b\dot{\phi}-\frac{1}{2}mlg\sin \phi }{\frac{1}{4}ml^2 + I},
\end{equation}
where $\phi$ is the pendulum angle measured anti-clockwise from the hanging down position, $g$ is the acceleration of gravity, $b$ is a friction coefficient and $I = \frac{1}{12}ml^2$ is the moment of inertia of a pendulum around the pendulum midpoint. In the experiment, we set $m=1 \mathrm{kg}, l=1\mathrm{m}, b=0.01$ and $g = 9.82\mathrm{m/s^2}$. We then define the system state as $x_{t_n}=[\dot{\phi}_{t_n},\phi_{t_n}]^\top$, and the extended control input as $v_n = [u_{t_n},\tau_n]^\top$. 
The cost functions are given by $c_1(\tau) = \tau_{\max} -\tau$, and $c_2(x) = -\exp(-\frac{1}{2}(\mathbf{Tri}(\phi)-\mathbf{Tri}(\pi))^\top Q (\mathbf{Tri}(\phi)-\mathbf{Tri}(\pi)))$,
where $\mathbf{Tri}(\cdot)$ is defined as $\mathbf{Tri}(\beta) = [\sin \beta, \cos \beta]^\top$ for $\beta \in \mathbb{R}$. Moreover, we set $Q = \mathrm{diag}(4,4)$, $\tau_{\min} = 0.02$, $\tau_{\max} = 0.6$ and $M=1$. 
The simulation result is given in Figs.\ref{pendulum_Sim}, \ref{pendulum_tau}. These results show that the proposed algorithm can learn an effective controller to stabilize the system towards the inverted position $\phi = -\pi$ (red dotted line). We can also see from Fig.\ref{pendulum_tau} that the inter-event time tends to be larger as $\lambda$ is selected larger, which is because we penalize more for the communication cost. 

For comparisons, we also conducted the experiment using \cite{learningbased}, in which \req{pendulumDyn} is approximated by the discrete-time system under the time period $0.02$. 
Table~\ref{simresulttable} shows the number of episodes and total execution time required to learn a controller that can stabilize the system to $\phi = -\pi$ within an error range 1\%. The table shows that the proposed approach achieves a significant reduction of the execution time to learn the controller, which is due to the fact that the previous approach \cite{learningbased} requires state-space discretization to solve the value iteration algorithm. Moreover, the proposed approach is more efficient, in the sense that it requires smaller episodes to learn the controller than \cite{learningbased}. This is because \cite{learningbased} is able to learn the dynamics only when the smallest inter-event time is selected, while the proposed approach can utilize all the data received at the controller to learn the dynamics during execution of the self-triggered controller.


\begin{table}[h]
\caption{Execution time and number of episodes to achieve stability}\label{simresulttable}
\begin{tabular}{ccc}
\hline
                            & \textbf{Previous approach \cite{learningbased}} & \textbf{Proposed approach} \\ \hline
\textbf{Execution time}     & 5678s                              & 96s                             \\
\textbf{Number of episodes} & 10                                 & 5                               \\ \hline
\end{tabular}
\end{table}

\subsection{Vehicle navigation with collision avoidance}
\label{Sim}
\begin{figure}
        \centering
        \includegraphics[width=2.35in]{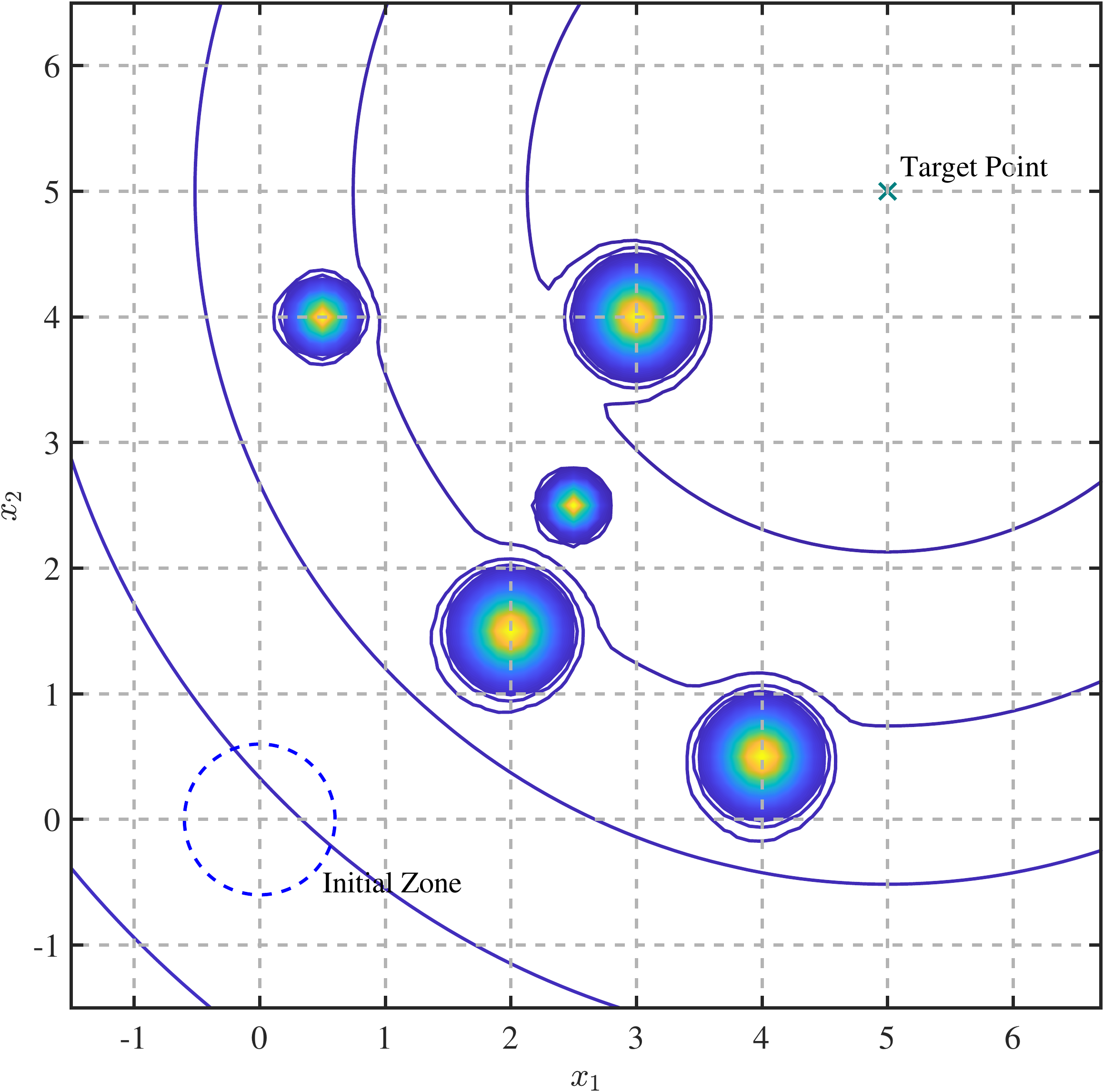}
        \caption{The contour map of $c_2(x)$}
        \label{cost}
\end{figure}

To test the benefits of incorporating the cost between the adjacent triggering instants as explained in Section~\ref{costfunction}, we next consider the following vehicle dynamical model: 
\begin{align}
  \dot{x}_1 = u_1 \cos(\theta),\    \dot{x}_2 = u_1 \sin(\theta),\  \dot{\theta} = u_2, 
\end{align}
where $[x_1,x_2]^\top \in \mathbb{R}^2$ is the two-dimensional coordinate of the vehicle, $\theta \in [0,2\pi)$ is the angle from the $x_1$-axis to the direction of the vehicle, $u_1$ is the velocity of the vehicle, and $u_2 \in [\omega_{\min},\omega_{\max}]$ is the angular velocity. 
We then define the system state as $x_{t_n} = [x_{1,t_n},x_{2,t_n},\theta_{t_n}]^\top$ and the system input as $v_n = [u_{1,t_n},u_{2,t_n},\tau_n]^\top$. 
The control goal is to drive the vehicle to a target position in a certain map where obstacles are placed, 
and it is assumed that the map and obstacle information is known. 
In addition, the vehicle is set to start from an initial zone and try to reach the target position. The stage cost $c_2$ is defined as 
\begin{equation*}
\label{defcost}
        \begin{split}
                c_2(x) &= c_{\mathrm{tg}}(x) + c_{\mathrm{ob}}(x), \\
                c_{\mathrm{tg}}(x) &= -K_{\mathrm{tg}}\exp\left(-\frac{1}{2}(x-x_{\mathrm{tg}})^\top Q(x-x_{\mathrm{tg}})\right),      \\
                c_{\mathrm{ob}}(x) &= K_{\mathrm{ob}}\sum_{\mathrm{ob} \in OB}\exp\left(-\frac{1}{2}(x-x_{\mathrm{ob}})^\top Q_{\mathrm{ob}}(x-x_{\mathrm{ob}})\right),   
        \end{split}
\end{equation*}
where $K_{\mathrm{ob}} > K_{\mathrm{tg}}>0$, $Q \succ 0$ is the weight matrix of the state $x$, $x_{\mathrm{tg}} \in \mathbb{R}^3$ is the target state,
$OB$ is the collection of all obstacles, and $x_{\mathrm{ob}} \in \mathbb{R}^3$, $Q_{\mathrm{ob}} \succ 0$ define the position and shape of obstacles respectively. 
In this simulation, we set $K_{\mathrm{tg}} = 1, K_{\mathrm{ob}} = 50, x_{\mathrm{tg}} = [5,5,0]^{T}$, and $Q = \mathrm{diag}(0.04,0.04,0)$. 
The contour map of $c_2(x)$ is shown in the Fig.\ref{cost}. As shown in the figure, we assume that there exist $5$ obstacles in the considered region.  
Moreover, the stage cost $c_1$ is given by $c_1(\tau) = \tau_{\max} - \tau$ with $\tau_{\max}= 0.8$, and the minimum inter-event time is set to $\tau_{\min} = 0.02$. 

\begin{figure}
        \centering
        \includegraphics[width=2.35in]{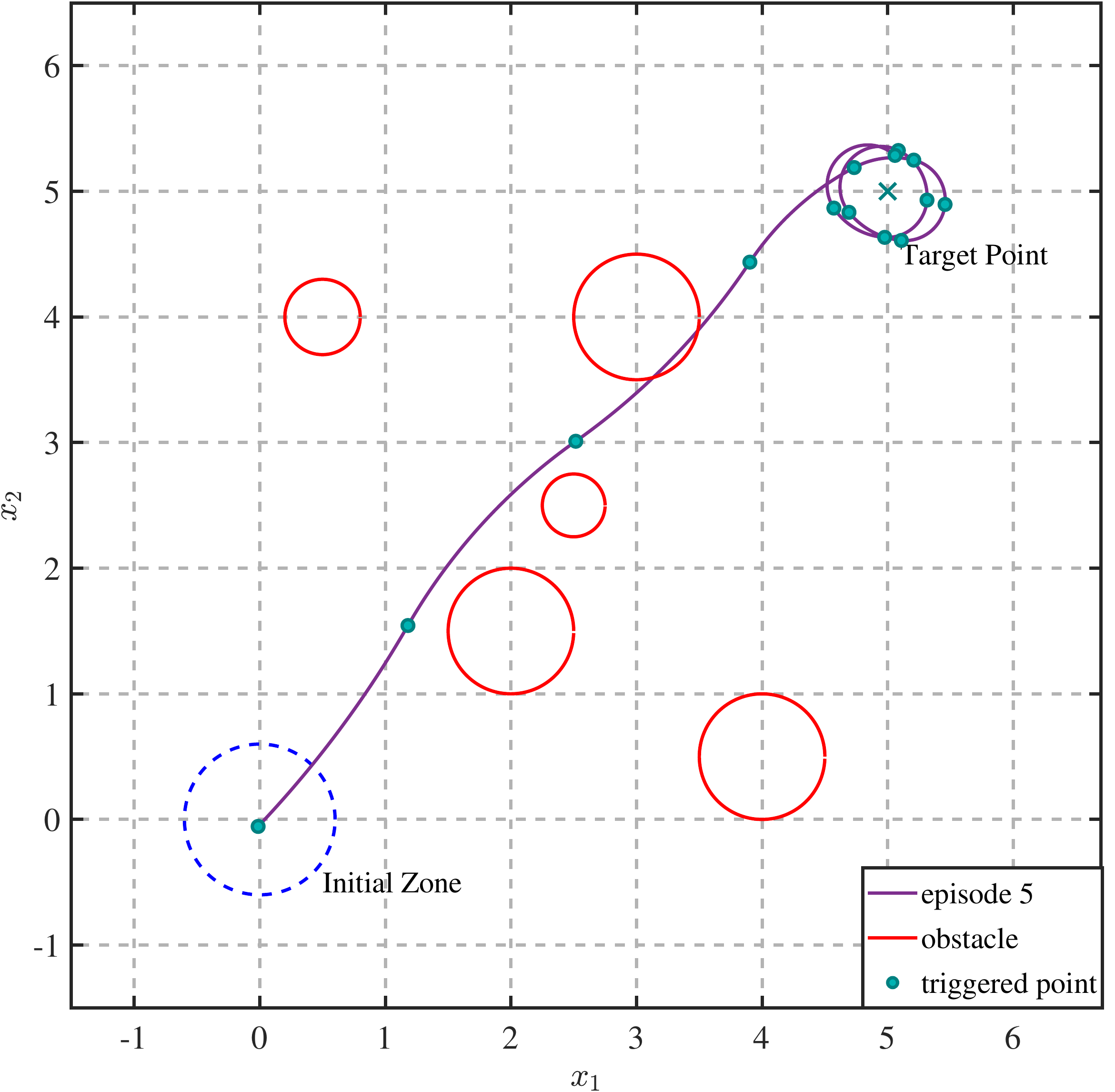}
        \caption{Learned controller performance when $M=1, \lambda=0.01$.}
        \label{M=1}
\end{figure}

\begin{figure}
        \centering
        \includegraphics[width=2.35in]{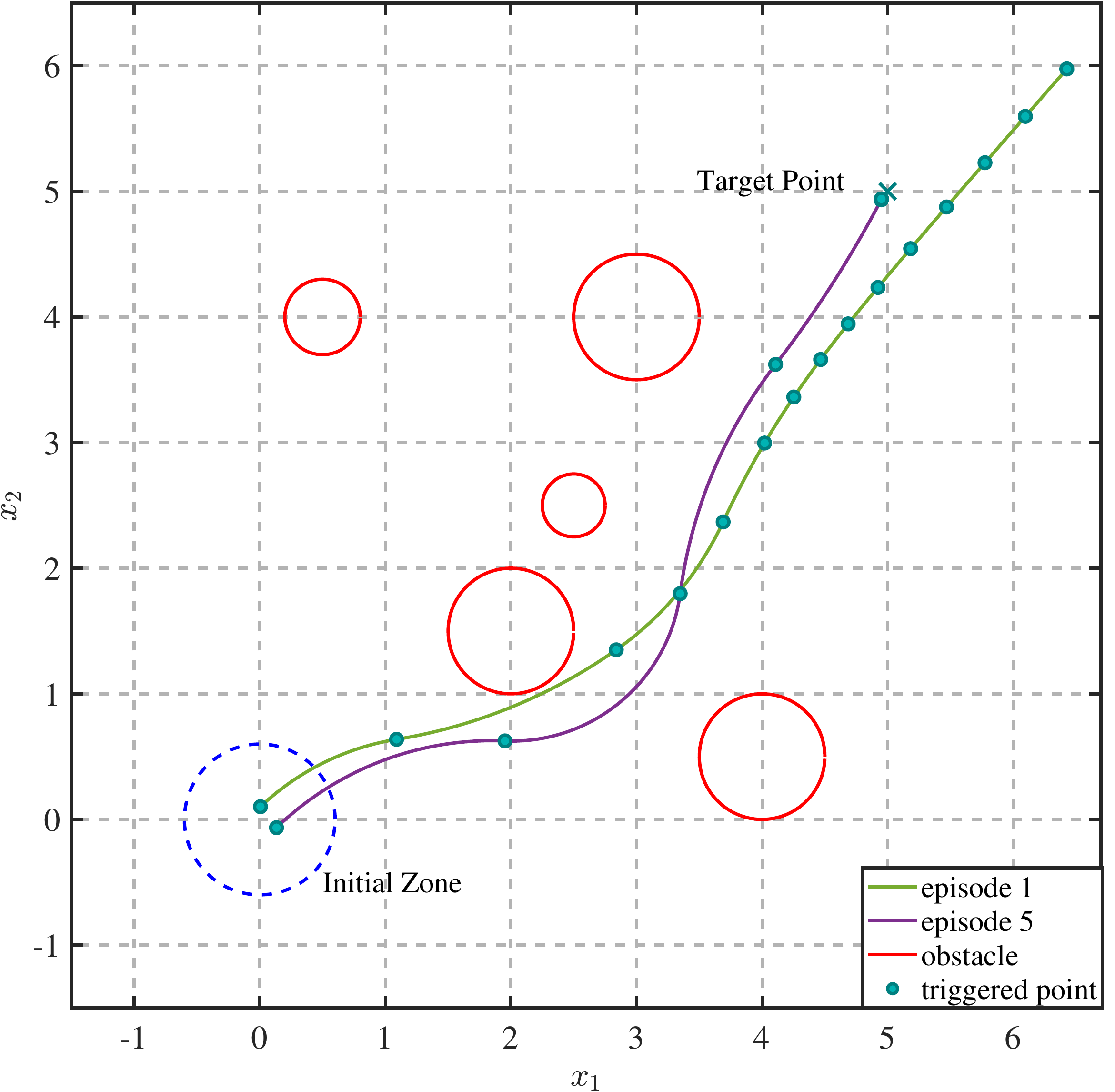}
        \caption{Learned controller performance when $M=5, \lambda=0.01$.}
        \label{M=5}
\end{figure}

\begin{figure}
        \centering
        \includegraphics[width=2.35in]{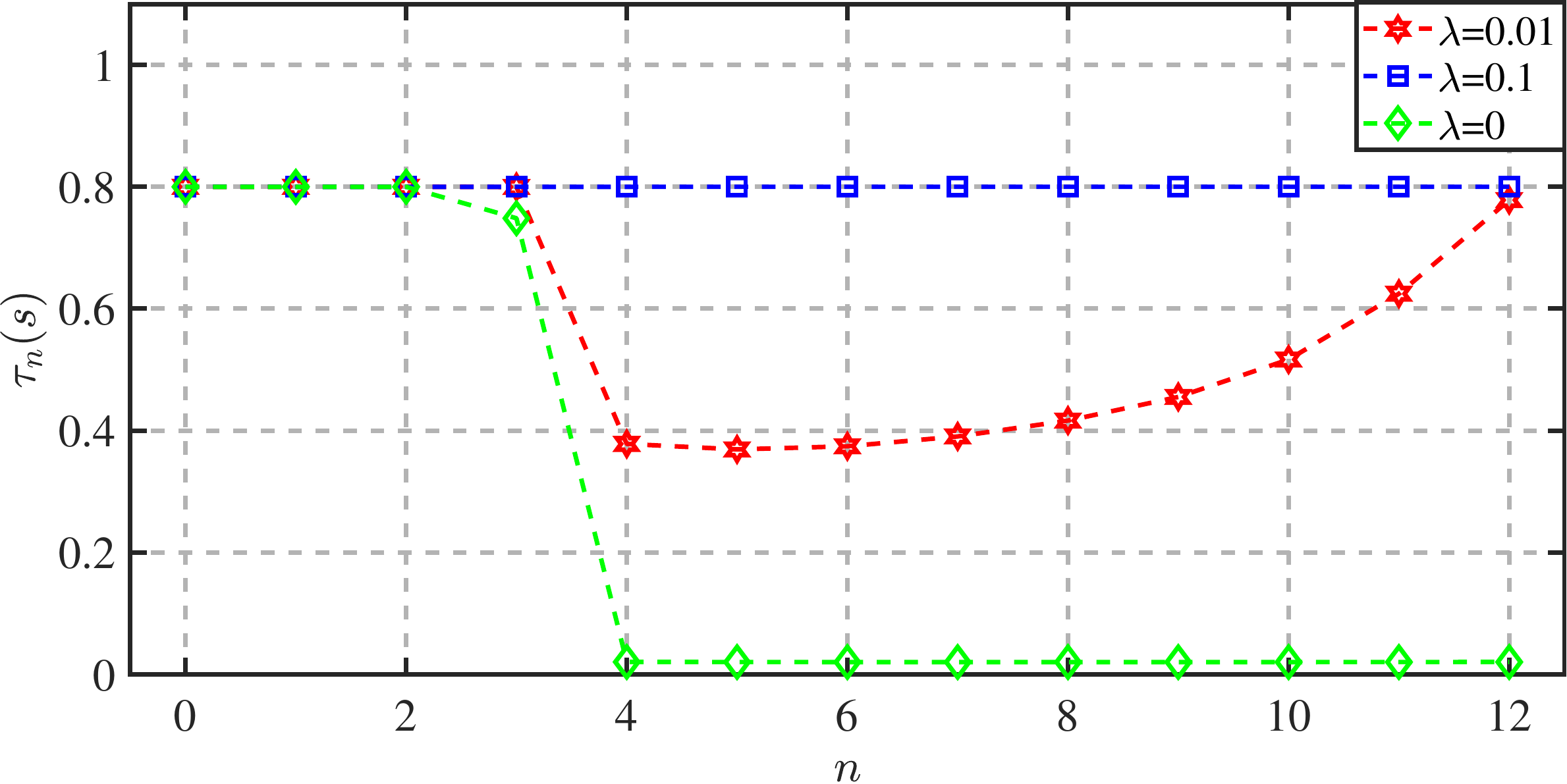}
        \caption{Inter-event time $\tau_n$ with $\lambda = 0.01,0.1,0$ when episode = $5$.}
        \label{tau}
\end{figure}

Performance of the controller learned from Algorithm~\ref{alg} is given in Figs.~\ref{M=1} and \ref{M=5}. 
It can be seen that when $M=1$, the vehicle directly goes through an obstacle (i.e., it fails to avoid an obstacle), 
which is attributed to the fact that when $M=1$, the expected total cost $J$ degenerates to \req{naivecost} and, as discussed in Section~III-D, it neglects state trajectories between any two adjacent triggering instants. 
Fig.\ref{M=5} describes the learning process, 
the learned system dynamics initially learns from random data, and at this time though the full dynamics has not been learned, 
it is enough to make the vehicle move roughly towards the target, 
then some unknown dynamics is detected, thus making the vehicle move more accurately. 
Fig.\ref{tau} depicts how inter-event times change with $\lambda=0.01,0.1,0$. 
It can be shown that regardless of $\lambda$, the learned controller tends to choose large $\tau_n$ at the beginning, 
which is mainly due to the fact that the total cost in (\ref{totalcost}) is defined by summing the stage costs only at the triggered instants and their interpolations. 
Reducing the number of triggering leads to the reduction of the total cost, 
and therefore, minimizing (\ref{totalcost}) leads to communication reduction even for the case $\lambda=0$ at the very beginning. 
However, when the vehicle is about to arrive at the target point, 
$\tau_n$ falls quickly for the case $\lambda = 0$, this is because the learned controller chooses not to tune the control inputs $u_1$ and $u_2$ but the inter-event time $\tau_n$ to minimize (\ref{totalcost}). 
Using a slightly larger $\lambda$ helps solve this problem, but a too large $\lambda$ causes the algorithm difficult to learn an effective controller.

\begin{figure}
        \centering
       \includegraphics[width=2.35in]{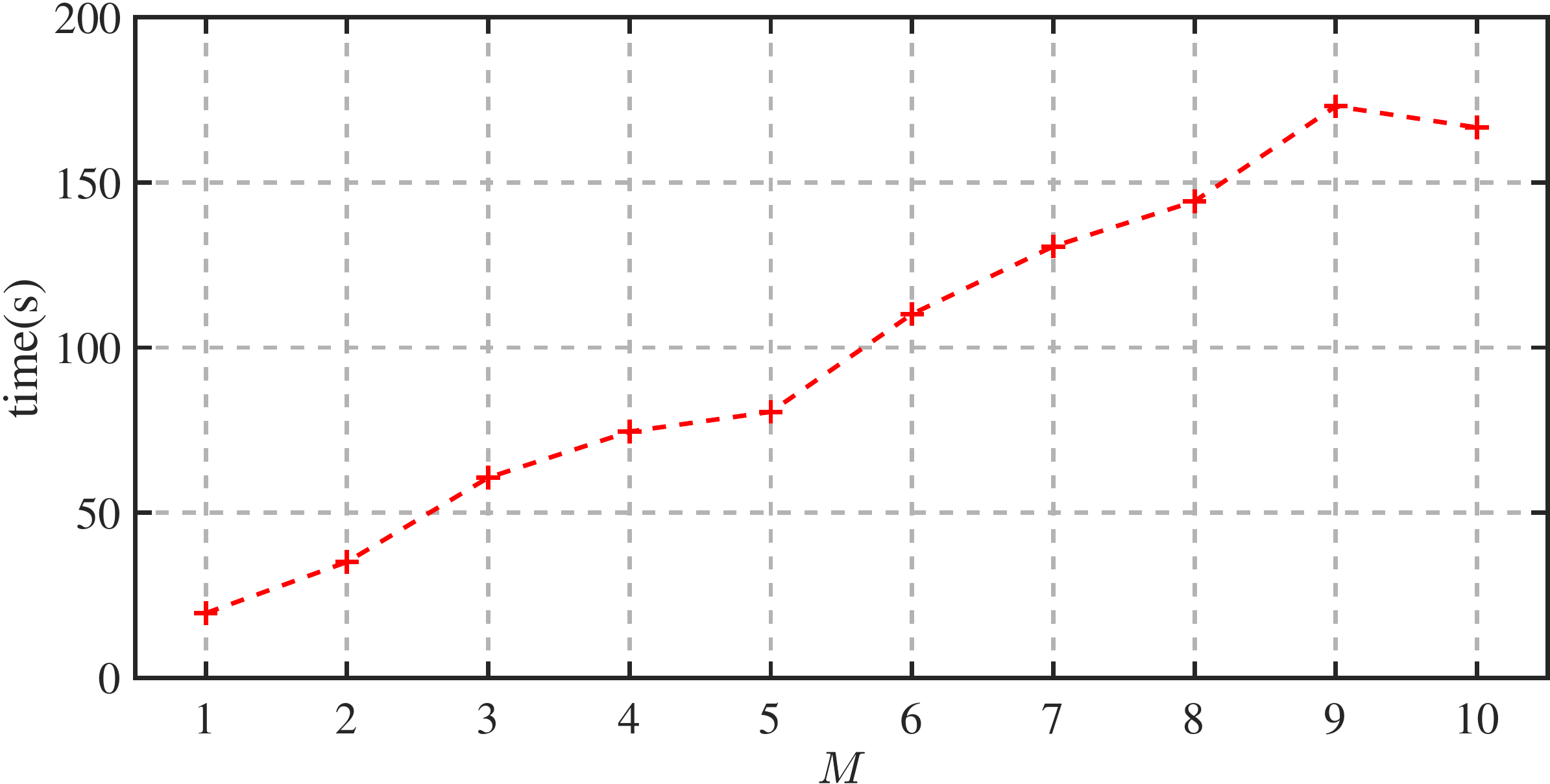}
        \caption{Runtime to take line~6-10 in Algorithm \ref{alg} with different $M$.}
        \label{runtime}
\end{figure}

Finally, we perform a complexity analysis of the proposed algorithm . The main complexity of the algorithm focuses at the policy improvement, namely lines 6-10 in Algorithm \ref{alg}. 
We use Fig.\ref{runtime} to show that how long running lines 6-10 takes with different $M$.  
It can be seen that the running time is approximately linear with $M$ which is attributed to the fact that increasing $M$ will only complexify the computation of $J$ and $\mathrm{d}J/\mathrm{d}\psi$ linearly.

\section{Conclusion}
In this paper, we studied the self-triggered control for NCSs with unknown transition dynamics. 
To this end, we lifted the original continuous dynamics to a novel discrete model by taking time as an input 
and used the GPR to learn the lifted dynamics of the plant. 
We formulated an optimal control problem where both the cost for the control performance and the communication cost are taken into account. 
Then, we illustrated that the minimization of the cost will produce an optimal self-triggered controller. 
In the simulation, detailed analysis of the simulation is given and shows that the proposed approach is effective and enjoys a low computational complexity.



\end{document}